\documentclass[11pt,twoside]{article}
\usepackage{asp2014}
\aspSuppressVolSlug
\resetcounters

\begin{document}

\title{On Thermohaline Mixing in Accreting White Dwarfs}
\author{Detlev~Koester
\affil{Institut f\"ur Theoretische Physik und Astrophysik,
  Universit\"at Kiel, Germany; \email{koester@astrophysik.uni-kiel.de}}
\paperauthor{Detlev~Koester}{koester@astrophysik.uni-kiel.de}{}{Universit\"at
Kiel}{Institut f\"ur Theoretische Physik und Astrophysik}{Kiel}{}{24098}{Germany}}

\begin{abstract}
We discuss the recent claim that the thermohaline (``fingering'')
instability is important in accreting white dwarfs, increasing the
derived accretion fluxes potentially by orders of magnitude. We present an
alternative view and conclude that at least in the steady state this
is not the case and the current method of estimating accretion fluxes
is correct.
\end{abstract}

\section{Introduction}
The thermohaline (saltfinger, fingering, double-diffusive) instability
is well-known in Oceanography: a warm layer of saltwater on top of a
cold body of freshwater may be dynamically stable but may nevertheless
lead to complete mixing, if the diffusion of heat is faster than that
of salt.  In the case of stars, the r\^ole of salt is played by the
molecular weight. A layer with higher molecular weight on top of a
layer with smaller weight may be dynamically stable (no convection),
but subject to a similar double-diffusive instability. Classical
papers in the astrophysical context are \cite{Ulrich72} and
\citet[][=KRT80]{Kippenhahn.Ruschenplatt.ea80}.

\section{The instability in the scenario of KRT80}
The instability starts from a boundary layer separating the two
layers. A front of a molecular weight gradient expands into the
homogenous region below, with decreasing velocity (proportional to the
decreasing gradient), leaving behind a (nearly) homogenous layer. This
can be described as a global mixing process for the average
concentration $\bar{c}$ of heavy elements given by a solution of the
type $\bar{c} \propto \exp(-t/\tau_{th})$.  This behavior is partly
analogous to a diffusion process, where an inhomogenous mixture
asymptotically approaches a homogenous state in the available volume
with a timescale $\tau_{th}$. KRT80 consequently define a diffusion
coefficient and the related timescale for a typical length $L$ of the
instable region as
\begin{equation}
D_{th} = C_{th}\,\frac{4 ac T^3}{3 c_p \kappa \rho^2} \,
        \left[\frac{-\nabla \mu}{\nabla_{ad}-\nabla}\right]
       = C_{th}\,\alpha\,\frac{1}{R_0} \mbox{\qquad and \qquad} 
       \tau_{th} \approx \frac{L^2}{D_{th}}.
\label{eqa}
\end{equation}
Here $C_{th}$ is a calibration constant, $\nabla$ are logarithmic
gradients, and the other symbols have their usual thermodynamic
meanings.  The whole discussion in KRT80 implies that the thermohaline
instability is a one-time event, which leads from an unstable
stratification to a homogenous mixture with a calculable timescale.
{\em KRT80 never define nor use a local diffusion velocity.}

\section{Application of the thermohaline instability in 
\citet[][=DDVV13]{Deal.Deheuvels.ea13}}  
We see several problems in the way the instability is applied and
treated in this paper.

\subsection{Local vs. global mixing, one-time event vs. continuous?}
Thermohaline instability is a one-time instability leading to a global
mixing with no well-defined local diffusion velocity. DDVV13 (and
earlier papers) extend this to a {\em continuous and local} process by
defining a local thermohaline diffusion velocity as
\begin{equation}
  v_{th} = D_{th} \, \frac{\partial \ln c}{\partial r} =
          -\frac{D_{th}}{H_p} \, \frac{\partial \ln c}{\partial \ln p}
\end{equation}
and combining this with continuous accretion and molecular diffusion.
Adding $v_{th}$ acting on the bulk material and the molecular
diffusion velocity $v_{12}$, which is a relative velocity between a
heavy atom and the main bulk material, different for each species, can
be regarded as a technical trick to facilitate computations. It is
meaningful only in the two limiting cases, where either of the two
processes is negligible. The analogy with convection, where this
method of combining ``convective mixing'' and diffusion is also
sometimes used, is misleading: the mixing coefficient in the
dynamically unstable convection zone is always orders of magnitude
larger than any molecular diffusion, and it is zero outside the cvz (+
overshooting region). A combination of the two mixing processes
therefore has never any practical importance.

\subsection{Definition of the thermohaline diffusion coefficient}
Widely differing coefficient have been used in the literature.  DDVV13
state that their coefficient is calibrated with numerical simulations
by \cite{Traxler.Garaud.ea11}. That paper defines a purely
empirical fit without physical basis, determined at Prandtl and Lewis
numbers of $\approx0.1$, whereas in the astrophysical context (see
below) these numbers are $\approx10^{-7} - 10^{-8}$, requiring an
extrapolation of a numerical fit over seven orders of magnitude.

The thermohaline instability applies only in a region where $\nabla <
\nabla_{ad} + \nabla_\mu$ (with $\nabla_\mu < 0$). This is the limit
of dynamical instability in case of a negative $\mu$ gradient (the
Ledoux criterion for convection). The expression in brackets in
eq.~\ref{eqa} ($=1/R_0$) goes to the limit 1, when approaching this
critical gradient. {\em In stark contrast to this behavior, DDVV13
  replace $1/R_0$ by an expression, which has an infinite singularity
  at the limit of the dynamically stable region}. This virtually
assures an instability at the boundary for arbitrary small $\mu$
gradients and cannot be physically sound. We therefore use the KRT80
formulation in our numerical estimates.

\section{Numerical example}
Although we have doubts in the validity of the DDVV13 approach, we
calculate the conditions at the bottom of the convection zone (cvz)
in the well-studied DA white dwarf G29-38 (11820~K, log g =8.4),
following their equations with the exception of the singularity in
$D_{th}$. Here we use the original KRT80 formulation.

The thermohaline diffusion velocity is $ v_{th} \propto c$. For small
number concentrations of heavy elements $c$, therefore $v_{th} \ll
v_{12}$, the molecular diffusion velocity, which is independent of
$c$. Likewise, the growth time of the linear instability is much
larger than the molecular diffusion timescale. In the beginning of an
accretion event therefore the diffusion equilibrium will be reached
before the instability develops. We have thus calculated the abundance
distribution for this equilibrium and determined the condition for
thermohaline mixing for this stratification.

Primary condition for instability is $R_0 < 1/Le$, which expresses the
condition that the excess heat is lost faster than the particles
diffuse. Only for the largest abundances is this condition fulfilled
in a small region below the cvz in G29-38, which has one of the
highest abundances observed in DAs.  For total metal abundances below
$10^{-7}$ the instability would never occur. Very similar conclusions
can be reached comparing the diffusion velocities (Fig.~1).

\articlefigure[width=1.0\textwidth]{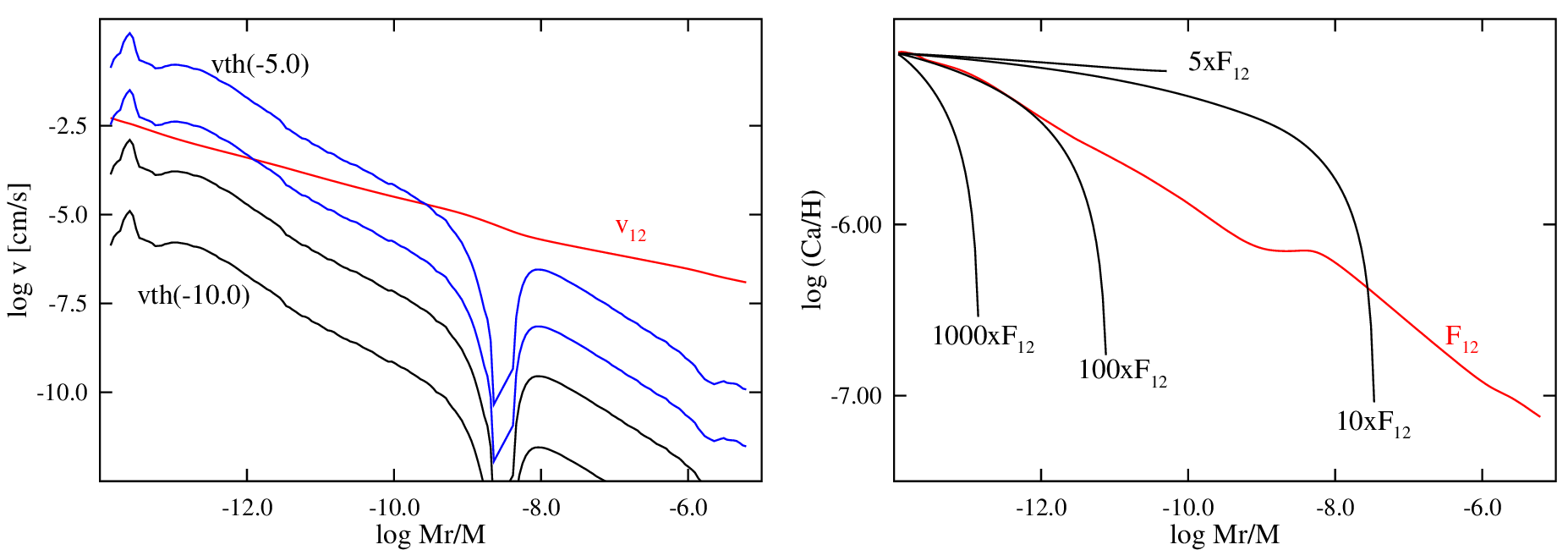}{fig}%
  {Left: molecular
  ($v_{12}$) and thermohaline ($vth$) diffusion velocities in the outer
  layers for various metal abundances log Ca/H from -5 to -10.  Right:
  Searching for steady state solutions for the heavy element abundance
  using only molecular diffusion ($F_{12}$) or including thermohaline
  diffusion with a flux increased by large factors.}

\section{No steady state solution with thermohaline mixing}
A steady state solution is characterized by a constant flux $F$ of
trace heavy particles at all layers. The sum of thermohaline mixing,
ordinary diffusion, and gravitational settling is
\begin{equation}
a_1 \left(\frac{dc}{d\ln p}\right)^2 + a_2 \left(\frac{dc}{d\ln
  p}\right) + \rho\, v_{12}\, c = F = \mbox{const}
\end{equation}
with coefficients $a_1, a_2$ independent of $c$.  We solve this
equation with the starting value $c$ at the bottom of the cvz and $F$
as a free parameter. For $F=\rho\, v_{12}\, c$ the standard steady
state solution with molecular diffusion alone is recovered. For larger
accretion fluxes the concentration gradient gets steeper with
increasing $F$, inevitably leading to $c=0$ or complex values
(Fig.~1). As a result {\em there is no steady state solution including
  thermohaline mixing}. We assume that the abundance in the cvz will
continue to rise until it reaches the ``standard'' value for molecular
diffusion alone, but this needs confirmation by time-dependent
calculations up to the final steady state.

\section{Astrophysical support}
Attempts to explain astrophysical phenomena (e.g. the abundance
patterns on the giant branches) have met with only limited success
\citep{Cantiello.Langer10,Wachlin.Miller-Bertolami.ea11,
  Theado.Vauclair12}, and often an increase of the mixing efficiency
by large factors, or invoking additional mixing processes, is required
to get agreement with observations.

The differences in maximum accretion rates between DA and DB white
dwarfs used by DDVV13 as confirmation of their calculations have a
much more plausible explanation in the uncertainties of the EOS for
liquid helium. On the other hand, a result from
\cite{Koester.Gansicke.ea14} lends support to our arguments.  We find
that the range of derived accretion rates in DA white dwarfs remains
the same between $10^{5.5}$ to $10^{8.5}$ g/s over the entire observed
range of $T_{eff}$ from 5000 to 25000~K, cooling times from 20\,Myr to
2\,Gyr, diffusion timescales from days to 100000\,yrs, from purely
radiative envelopes to very deep convection zones, and observed Ca or
Si abundances from $10^{-12}$ to $10^{-4.5}$. It is highly unlikely
that such a consistent result would be achieved with the inclusion of
a thermohaline mixing description as in DDVV13.

\section{Conclusions}
Thermohaline instability may play an important r\^ole in
astrophysics. Possible examples are the cases discussed in KRT80 with
differences of the molecular weight of the order of 1, or catastrophic
events like the sudden infall of a 0.03\,$M_{jup}$ object on a star
(Theado \& Vauclair 2012). Concerning the accretion on white dwarfs,
with a gradual build-up of heavy element abundances and $\mu$
differences of $10^{-6}$ or smaller, the situation is quite different.

The validity of extending the instability to a continuous process, the
mixing with molecular diffusion, and the extrapolation of mixing
efficiencies over seven orders of magnitude in Lewis numbers, is not
obvious. The constancy of derived accretion fluxes in DAs over an
extreme range of all parameters is a strong argument against the
importance of the thermohaline instability in this scenario.

A more detailed version of this study with more figures can be found at\\
{\small
\url{www1.astrophysik.uni-kiel.de/~koester/astrophysics/astrophysics.html}
}

\end{document}